\documentclass[conference]{IEEEtran}
\PassOptionsToPackage{table}{xcolor}
\usepackage{graphicx}
\usepackage{multirow}
\usepackage{amsmath}
\usepackage{amssymb,soul}
\usepackage{xcolor}
\usepackage{threeparttable}
\usepackage{booktabs}
\usepackage{colortbl}
\usepackage{bbm}
\usepackage{algorithm}
\usepackage{algpseudocode}

\definecolor{rowgray}{gray}{0.92}

\ifCLASSINFOpdf

\else

\fi

\begin{document}

\title{\LARGE{Temporal and Multimodal Deep Learning for Cyberattack Detection in LEO Satellite Systems\\
}}

\author{
\IEEEauthorblockN{
Kyle Stein,
Guillermo Francia, III,
Eman El-Sheikh,
Hossain Shahriar
}
\IEEEauthorblockA{
Center for Cybersecurity and AI\\
University of West Florida\\
Pensacola, FL, USA\\
\{kstein, gfranciaiii, eelsheikh, hshahriar\}@uwf.edu
}
}
\maketitle

\begin{abstract}
The growing reliance on Low-Earth Orbit (LEO) satellite communication systems has increased the need for intelligent methods capable of detecting cyberattacks across complex and dynamic space environments. Unlike conventional network intrusion detection, satellite systems generate heterogeneous information across radio-frequency (RF) links, onboard hardware, and orbital operations. However, many existing approaches either rely on terrestrial intrusion datasets or evaluate individual observations independently, limiting their ability to capture temporal attack behavior specific to LEO satellites. In this work, we conduct a systematic study of deep-learning-based cyberattack detection using the recently introduced satellite-specific UNSW-IoTSAT dataset. We investigate structured learning architectures that preserve hardware, orbital, and RF information, including a Subsystem-Fusion MLP and a hierarchical multimodal Transformer that models both cross-subsystem interactions and temporal evolution. We further evaluate leakage-resistant row-level and temporal settings, along with cross-satellite generalization, to characterize how model architecture and evaluation protocol influence satellite cyberattack detection. Experimental results demonstrate the value of structured multimodal modeling and rigorous evaluation, with the hierarchical Transformer achieving up to 91.66\% accuracy and 85.63\% macro F1 under the leakage-resistant evaluation protocol.
\end{abstract}


\vspace{-1mm}
\section{Introduction}
\label{sec:introduction}
\begin{figure}[t!]
    \centering
    \includegraphics[width=0.9\columnwidth]{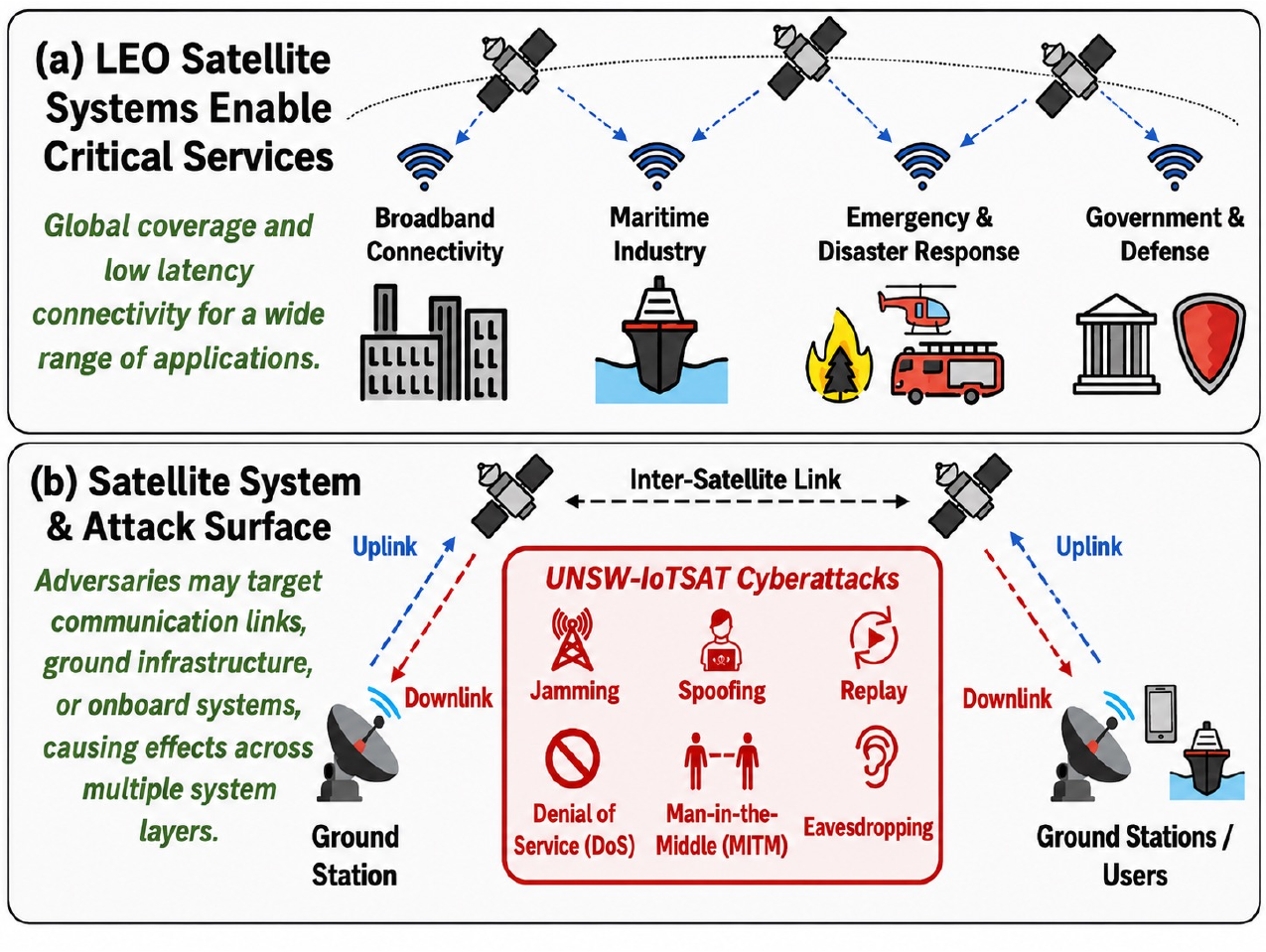}
    \caption{\textbf{LEO satellite services and cyberattack surface. (a) Critical services supported by LEO satellite systems. (b) Satellite communication links and cyberattacks represented in the UNSW-IoTSAT dataset.}}
    \label{fig:intro}
    \vspace{-5mm}
\end{figure}

Space-based infrastructure has become an increasingly important component of modern communication systems \cite{kodheli2020satellite}. Low-Earth Orbit (LEO) satellite communication networks now support a wide range of services, including: broadband connectivity, Internet of Things (IoT) applications, maritime operations, emergency and disaster response, and government and defense services \cite{prol2022position,yue2023low}. As dependence on satellite-enabled services continues to grow, disruptions to these systems have serious cascading consequences. This risk was demonstrated in the 2022 Viasat KA-SAT cyberattack, where attackers compromised ground infrastructure used to manage satellite terminals and rendered tens of thousands of customer modems inoperable across Europe \cite{boschetti2022space}. Incidents such as these highlight the urgent need for intelligent systems capable of detecting malicious activity within LEO satellite communication systems. Figure \ref{fig:intro} provides an overview of this context, illustrating (a) critical services enabled by LEO satellite systems and (b) the communication architecture and cyberattack surface.

However, securing LEO satellite communications from adversarial threats involves more than protecting the contents of communication links. Satellite operations rely on complex interactions among communication hardware, onboard systems, orbital state, and radio-frequency (RF) links. Consequently, malicious activity may produce observable effects across multiple sources of system information. For example, RF measurements may capture changes associated with communication-layer attacks, while hardware and orbital telemetry can provide complementary information about the state of the satellite system. Therefore, AI-enabled intrusion detection systems (IDSs) that jointly analyze these heterogeneous measurements may provide a more complete representation of malicious behavior than approaches based on a single source of information. 

Recent advances in deep learning have transformed cybersecurity by enabling models to identify complex patterns of malicious behavior across large volumes of system and network data \cite{stein2024revolutionizing,neto2025deep}. These capabilities are promising for satellite security since learning-based IDSs can model relationships among multiple measurements rather than relying solely on predefined signatures or isolated indicators. However, progress has been constrained by the limited availability of public datasets that support \textit{multiclass cyberattack} analysis while \textit{jointly representing} LEO satellite operational and communication behavior. Consequently, much of the existing deep learning literature has focused on individual communication-layer threats, such as spoofing and jamming \cite{wigchert2025detection,mehr2024deep}. The recently introduced UNSW-IoTSAT dataset \cite{ABDELHAMEED2026100133, abdelhameed2026unswiotsat} provides an important foundation for broader evaluation by combining labeled hardware, orbital, and RF measurements under benign and multiclass adversarial conditions.

Although this dataset enables broader evaluation of satellite cyberattack detection, several questions remain regarding how such data should be modeled and evaluated. First, satellite telemetry is inherently temporal, as each individual measurement captures only the system state at a single point in time, while cyberattacks may evolve across a sequence of observations. Treating each observation independently at the row level may therefore discard important information contained in the recent behavior of the system. Second, evaluation methodology can significantly influence reported detection performance. Consecutive observations generated during the same attack occurrence are often highly related, and randomly distributing these observations across train and test sets may lead to strong generalization on paper, but fail to reflect in real-world operational environments. Similar concerns arise when observations from the same satellite appear throughout training and testing, since a model may partially rely on satellite-specific characteristics rather than learning attack behavior that generalizes across different satellite systems.

To address these challenges, this work conducts a systematic study of row-level, temporal, and cross-satellite deep-learning based cyberattack detection on the UNSW-IoTSAT dataset. We develop two structured architectures, a Subsystem-Fusion MLP and a hierarchical multimodal Transformer, that preserve hardware, orbital, and RF information while modeling their relationships within and across satellite observations. We further introduce leakage-resistant data partitioning, matched row-level and temporal evaluation, and bidirectional cross-satellite testing to examine how model architecture, temporal context, and satellite-specific characteristics influence cyberattack detection and generalization. Overall, our key contributions can be summarized as follows:

\begin{itemize}
    \item We propose both a subsystem-fusion MLP and hierarchical multimodal Transformer that jointly model hardware/environmental, orbital/kinematic, and RF telemetry across consecutive satellite observations for multiclass cyberattack detection.
    \item We introduce a leakage-resistant evaluation framework and a matched row-level versus temporal comparison to quantify whether temporal context provides meaningful detection gains beyond individual observations.
    \item We evaluate cross-satellite generalization by training on one satellite and testing on a completely unseen satellite, assessing whether learned representations capture transferable cyberattack behavior. 
\end{itemize}

\section{Related Work}
\label{sec:related-work}
LEO satellite communication systems consist of interconnected spacecraft, communication links, ground infrastructure, and user terminals, creating attack surfaces that extend across both cyber and physical components \cite{sharmin2025cyber,enisa2024leo}. Recent studies have characterized these threats from both adversarial and system-level perspectives. Peled et al. \cite{peled2023evaluating}, for example, extend concepts from the MITRE ATT\&CK framework to describe adversarial behavior throughout the satellite attack lifecycle, while Willbold et al. \cite{willbold2023space} demonstrate practical vulnerabilities in onboard firmware and telecommand interfaces. Other studies identify potential attack surfaces across individual spacecraft subsystems, including attitude determination and control, onboard computing, communications, power, and mission payloads \cite{verma2025cybersecurity,salim2024cybersecurity}. Therefore, malicious activity affecting one component may also produce observable changes elsewhere in the cyber-physical satellite system.

\begin{table*}[t!]
\centering
\caption{Satellite feature groups used for attack detection.}
\label{tab:feature_groups}
\small
\setlength{\tabcolsep}{4pt}
\begin{tabular}{@{}lcp{11.0cm}@{}}
\toprule
\textbf{Feature Group} & \textbf{Count} & \textbf{Measurements} \\
\midrule
Hardware/Environmental & 12 &
Shunt voltage, current, power, three-axis magnetic field, proximity, ambient light, magnetic magnitude, power density, light-to-proximity ratio, and power-to-magnetic-field ratio. \\

Orbit/Kinematic & 9 &
Latitude, longitude, altitude, north/east/up velocity, total speed, distance from the reference origin, and velocity bearing. \\

Radio Frequency & 10 &
CRC errors, synchronization-word detections, received signal strength, SNR, bit-error rate, packet-error rate, throughput, frequency offset, Doppler shift, and constellation error. \\
\midrule
\textbf{Total} & \textbf{31} & \\
\bottomrule
\end{tabular}
\end{table*}

Consequently, machine and deep learning have been investigated for automated detection of malicious attacks in satellite systems. Conventional approaches include support vector machines for satellite interference recognition \cite{yang2019efficient} and decision-tree intrusion detection for integrated satellite-terrestrial networks \cite{yap2022network}. Artificial neural-network feature extraction has also demonstrated promising intrusion-detection performance \cite{koroniotis2022new,azar2023deep}. These studies demonstrate the potential of learned representations for satellite cybersecurity; however, many existing approaches focus on a specific communication threat, such as spoofing or jamming, or are evaluated using conventional network and IoT datasets that provide limited representation of satellite-specific cyber-physical behavior. Moreover, satellite telemetry is inherently sequential, and comparatively limited attention has been given to determining whether explicitly modeling consecutive observations improves attack detection over row-wise classifiers.

Dataset availability further limits the evaluation of satellite intrusion-detection methods. Operational spacecraft datasets such as the ESA anomaly-detection benchmark \cite{kotowski2024european} and OPS-SAT \cite{ruszczak2025ops} provide valuable real-world telemetry, but primarily characterize operational anomalies rather than labeled cyberattacks, which is necessary for proper machine/deep learning experimentation. Furthermore, evaluation on closely related observations from the same simulated environment makes it difficult to determine whether a detector has learned transferable attack characteristics across satellite systems. In particular, consecutive observations belonging to the same attack event may be highly similar, and evaluation on the same satellite used during model development does not directly measure generalization to a different satellite platform.

UNSW-IoTSAT \cite{ABDELHAMEED2026100133, abdelhameed2026unswiotsat} provides a unique opportunity to investigate these limitations by combining hardware, orbital, and RF measurements with labeled normal behavior and multiclass cyberattack categories across two satellite nodes. Rather than introducing another attack-specific detector, this work leverages UNSW-IoTSAT to systematically examine how model architecture, temporal context, and evaluation setting influence satellite multiclass cyberattack detection. We compare machine-learning/deep-learning models under leakage-resistant attack-instance partitions, evaluate row-level and temporal models on matched prediction targets, and measure cross-satellite generalization when the target satellite is excluded from model training.

\section{Dataset and Preprocessing}
\label{sec:data}
The UNSW-IoTSAT dataset was developed to support cybersecurity research for IoT-enabled satellite communication systems. The dataset was generated using a hybrid satellite IoT testbed that combines real onboard sensor measurements, simulated orbital trajectories, and RF communication between satellite nodes and a ground station. The dataset captures normal operation and six cyberattack categories: jamming, spoofing, replay, denial-of-service (DoS), man-in-the-middle (MITM), and eavesdropping. Each observation contains measurements describing hardware/environmental conditions, orbital/kinematic behavior, and RF/communication characteristics. Furthermore, UNSW-IoTSAT contains observations from two distinct satellite nodes, Satellite1 and Satellite2, enabling evaluation of whether learned attack-detection models generalize across satellite systems.

\vspace{1mm}
\noindent\textbf{Dataset Cleaning and Satellite Features:}
We preserve the original ordering of the observations and normalize the attack labels into seven target classes: normal, jamming, spoofing, DoS, MITM, replay, and eavesdropping. Rather than automatically retaining all numerical fields provided by the dataset, we define an explicit 31-feature list representing three complementary sources of satellite information. The hardware features characterize power consumption, magnetic-field behavior, and environmental sensing. The orbital features describe simulated position and motion. Lastly, the RF features characterize communication quality and receiver behavior. The three feature groups used throughout this study are summarized in Table \ref{tab:feature_groups}. Timestamps, satellite identifiers, source-row identifiers, attack-instance identifiers, and attack-related metadata are retained only for partitioning and auditing and are never supplied to the classifiers. This restricts the models to learn representations from the measurements of underlying physical, orbital, and communication state.

\vspace{1mm}
\noindent\textbf{Leakage-Resistant Data Partitioning:}
Consecutive observations generated during the same cyberattack may be highly dependent, and placing samples from the same attack occurrence in both training and testing can introduce data leakage and inflate estimates of generalization \cite{flood2024bad}. To reduce this form of leakage, malicious observations are grouped according to their attack timeline, while normal observations are organized into contiguous operating blocks independently for each satellite. Complete attack instances and normal blocks are assigned to a single partition and are never divided across training and testing. For temporal evaluation, these partitions are fixed before any temporal windows are constructed. Sliding windows are then generated independently within each partition and satellite, ensuring that observations contained in training windows cannot also appear in test windows. For cross-satellite evaluation, one satellite instead serves as the source and training domain and the other as a completely held-out testing domain. Both Satellite1-to-Satellite2 and Satellite2-to-Satellite1 transfer directions are evaluated.

\begin{table}[t!]
\centering
\caption{Class distribution of matched temporal-window targets for the primary evaluation protocol.}
\label{tab:temporal_distribution}
\footnotesize
\setlength{\tabcolsep}{7pt}
\begin{tabular}{@{}lrr@{}}
\toprule
\textbf{Class} & \textbf{Train} & \textbf{Test} \\
\midrule
Normal         & 129,670 & 33,286 \\
Jamming        &   8,780 &  2,277 \\
Spoofing       &  51,616 &  7,064 \\
DoS            &  13,592 &  4,505 \\
MITM           &   6,069 &  2,893 \\
Replay         &   1,652 &    345 \\
Eavesdropping  &  47,954 & 16,185 \\
\midrule
\textbf{Total} & \textbf{259,333} & \textbf{66,555} \\
\bottomrule
\end{tabular}
\end{table}

\begin{figure*}[t!]
    \centering
    \includegraphics[width=0.85\textwidth]{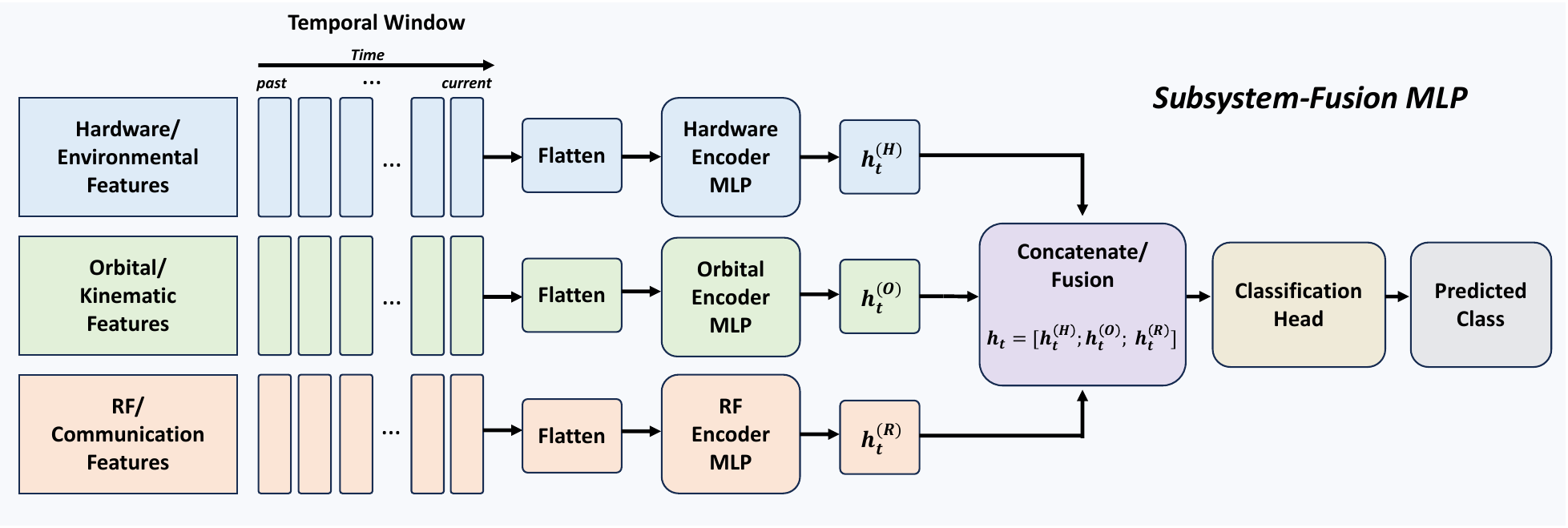}
    \caption{\textbf{Overview of the Subsystem-Fusion MLP cyberattack detection framework.}}
    \label{fig:method_mlp}
\end{figure*}

\vspace{1mm}
\noindent\textbf{Temporal Window Construction:}
Temporal examples are constructed only after the data partitions have been fixed. Within each partition, observations are ordered independently for each satellite according to their timestamps, with the original source-row index used to resolve ties. For a temporal window of length $L$ ending at observation $t$:

\begin{equation}
\mathcal{W}_{t}
=
\left[
\widetilde{\mathbf{x}}_{t-L+1},
\ldots,
\widetilde{\mathbf{x}}_{t}
\right]
\in \mathbb{R}^{L \times D},
\label{eq:temporal_window}
\end{equation}

\noindent where $\widetilde{\mathbf{x}}_{t} \in \mathbb{R}^{D}$ denotes the preprocessed telemetry vector at observation $t$, and $D$ denotes the number of retained input features. Temporal windows are constructed with a stride of one, and the class of the final observation is used as the prediction target. Labels from earlier observations are not supplied to the model, and windows may contain transitions between normal and malicious behavior or between different attack categories. This continuous-timeline construction allows the models to observe how satellite behavior evolves around attack transitions rather than restricting temporal sequences to class-pure observations.

The resulting dataset is naturally imbalanced, with normal operation and several attack categories occurring substantially more frequently than others. We preserve this distribution rather than artificially balancing the training or test sets so that evaluation reflects the underlying dataset composition. Table \ref{tab:temporal_distribution} summarizes the class distribution of the matched temporal training and test targets. The \textit{same} prediction endpoints are used for the corresponding row-level experiments, allowing temporal and non-temporal models to be compared on identical target observations.

\vspace{-1mm}
\section{Proposed Method}
\label{sec:proposed-method}
In this section, we describe the model architectures and training procedure used to detect benign and malicious behavior in LEO satellite systems. We investigate two structured deep-learning models: a subsystem-fusion MLP and a hierarchical multimodal Transformer. These two architectures constitute the primary methodological implementations of this work. Both proposed architectures explicitly organize the input according to hardware, orbital, and RF measurements. The subsystem-fusion MLP processes these feature groups through separate modality-specific encoders before combining their learned representations for classification. The hierarchical Transformer instead uses self-attention \cite{vaswani2017attention} to model interactions among the three satellite subsystems within each observation and then applies a second Transformer stage to model how the resulting satellite-state representations evolve over time. Therefore, the two architectures provide complementary approaches for evaluating the importance of subsystem information and temporal context in satellite cyberattack detection.

\begin{figure*}[t!]
    \centering
    \includegraphics[width=1.0\textwidth]{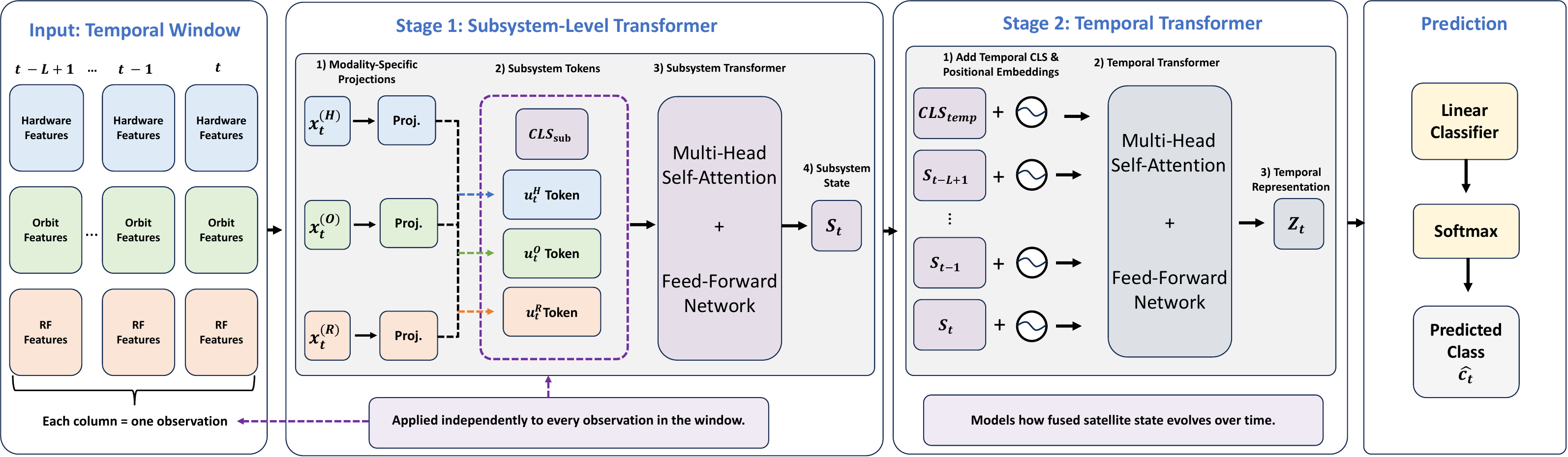}
    \caption{\textbf{Overview of the Hierarchical Multimodal Transformer cyberattack detection framework.}}
    \label{fig:method_transformer}
\end{figure*}

\subsection{Subsystem-Fusion MLP}
\label{sec:subsystem_mlp}

We first develop a Subsystem-Fusion MLP that preserves the three satellite information sources before classification. Rather than directly concatenating all telemetry measurements into a single input vector, hardware, orbital, and RF measurements are processed independently. This allows each encoder to learn  representations specialized to the characteristics of its corresponding satellite subsystem. An overview of the proposed subsystem-fusion MLP is shown in Fig.~\ref{fig:method_mlp}


For observation $t$, let $\mathbf{x}^{(m)}_t \in \mathbb{R}^{D_m}$ denote the modality-specific subset of $\widetilde{\mathbf{x}}_t$, where $m\in\{H,O,R\}$ represents hardware, orbital, and RF information. For a temporal window of length $L$, let $\mathbf{X}^{(m)}_t \in \mathbb{R}^{L\times D_m}$ contain the corresponding measurements. Each modality is encoded as:

\begin{equation}
\mathbf{h}^{(m)}_t =
f_m\!\left(\operatorname{vec}\!\left(\mathbf{X}^{(m)}_t\right)\right),
\label{eq:mlp_subsystem}
\end{equation}

\noindent where $f_m(\cdot)$ is a modality-specific MLP encoder and $\operatorname{vec}(\cdot)$ flattens the temporal measurements while preserving their chronological ordering. The resulting subsystem representations are concatenated to form:

\vspace{-2mm}
\begin{equation}
\mathbf{h}_t =
\left[
\mathbf{h}^{(H)}_t;
\mathbf{h}^{(O)}_t;
\mathbf{h}^{(R)}_t
\right].
\label{eq:mlp_fusion}
\end{equation}

The fused representation is then supplied to a classification network $g_{\mathrm{cls}}(\cdot)$ which maps the concatenated subsystem representation to logits over the target classes:

\vspace{-2mm}
\begin{equation}
\hat{\mathbf{y}}_t =
g_{\mathrm{cls}}(\mathbf{h}_t),
\qquad
g_{\mathrm{cls}}:\mathbb{R}^{d_f}\rightarrow\mathbb{R}^{C},
\label{eq:mlp_classifier}
\end{equation}

\noindent where $d_f$ denotes the dimensionality of the fused representation $\mathbf{h}_t$, and $C$ is the number of target classes. This architecture provides a useful complementary comparison to the proposed Transformer since both models explicitly preserve the hardware, orbital, and RF structure of the satellite measurements. However, the Subsystem-Fusion MLP combines fixed subsystem representations through concatenation and does not explicitly learn attention-based relationships among modalities or across individual satellite states.

\subsection{Hierarchical Multimodal Transformer}
\label{sec:hierarchical_transformer}
The proposed hierarchical multimodal Transformer is designed to capture two
complementary forms of structure in satellite telemetry. First, each observation contains measurements describing different parts of the satellite system, including hardware and environmental conditions, orbital and kinematic state, and RF communication behavior. Second, satellite behavior evolves over time, meaning that the significance of the current observation may depend on how the system state has changed over the preceding observations. Therefore, the architecture operates in two stages: a subsystem-level Transformer that models subsystem relationships within each observation and a temporal Transformer that models relationships across consecutive satellite states. An overview of this architecture is displayed in Figure \ref{fig:method_transformer}.

\vspace{1mm}
\noindent\textbf{Stage 1: Subsystem-Level Transformer.}
For observation $t$, let $\mathbf{x}^{(H)}_t$,
$\mathbf{x}^{(O)}_t$, and $\mathbf{x}^{(R)}_t$ denote the
hardware/environmental, orbital/kinematic, and RF feature vectors, respectively. Since the three groups contain different types and values of measurements, each is independently mapped to a common $d$-dimensional embedding space using a learnable modality-specific linear projection:

\vspace{-3mm}
\begin{equation}
\small
\mathbf{u}^{H}_t=\phi_H(\mathbf{x}^{(H)}_t), \qquad
\mathbf{u}^{O}_t=\phi_O(\mathbf{x}^{(O)}_t), \qquad
\mathbf{u}^{R}_t=\phi_R(\mathbf{x}^{(R)}_t),
\label{eq:modality_projection}
\end{equation}
\vspace{-3mm}

\noindent where $\phi_H$, $\phi_O$, and $\phi_R$ denote the learnable projection functions for the three satellite information sources. The resulting vectors $\mathbf{u}^{H}_t$, $\mathbf{u}^{O}_t$, and $\mathbf{u}^{R}_t$ form compact representations of the hardware, orbital, and RF conditions observed at the same point in time and are treated as subsystem tokens. A learnable subsystem classification token $\mathrm{CLS}_{\mathrm{sub}}$ is prepended to these tokens to form the subsystem token sequence $\left[\mathrm{CLS}_{\mathrm{sub}};\mathbf{u}^{H}_t; \mathbf{u}^{O}_t;\mathbf{u}^{R}_t\right]$. The resulting sequence contains four tokens: one learnable classification token and one token representing each satellite information source, preserving the subsystem organization of the telemetry while allowing the model to learn relationships among hardware, orbital, and RF conditions within the same observation. 

Self-attention is then applied across the tokens using the standard scaled dot-product formulation:

\vspace{-2mm}
\begin{equation}
\operatorname{Attn}(\mathbf{Q},\mathbf{K},\mathbf{V})
=
\operatorname{softmax}
\left(
\frac{\mathbf{Q}\mathbf{K}^{T}}{\sqrt{d_k}}
\right)
\mathbf{V},
\label{eq:attention}
\end{equation}
\vspace{-2mm}

\noindent where $\mathbf{Q}$, $\mathbf{K}$, and $\mathbf{V}$ denote the query, key, and value representations obtained from learned projections of the input tokens, and $d_k$ denotes the dimensionality of the key representations. Multi-head self-attention allows these relationships to be modeled across multiple representation subspaces. In the satellite setting, this enables an RF condition, for example, to be interpreted jointly with the corresponding hardware and orbital conditions rather than independently. After the subsystem tokens interact through self-attention, the output corresponding to $\mathrm{CLS}_{\mathrm{sub}}$ is retained as $\mathbf{s}_t$, the fused representation of the satellite state at observation $t$. Therefore, $\mathbf{s}_t$ summarizes information from all three satellite subsystems and serves as the input representation for the temporal modeling stage.

\vspace{1mm}
\noindent\textbf{Stage 2: Temporal Transformer.}
The subsystem-level Transformer from Stage 1 is applied independently to each observation in a temporal window of length $L$. This produces $L$ integrated satellite-state representations, $\mathbf{s}_{t-L+1},\ldots,\mathbf{s}_{t}$, where each $\mathbf{s}_t$ summarizes the combined hardware, orbital, and RF state at a single observation. Unlike the subsystem tokens used in Stage 1, these representations describe the complete satellite state rather than an individual information source. Now, Stage 2 of this architecture models how the complete satellite state changes across consecutive observations.

A learnable temporal classification token $\mathrm{CLS}_{\mathrm{temp}}$ is prepended to the sequence of satellite-state representations. Learnable positional embeddings $\mathbf{P}$ are also added to preserve the chronological ordering of the observations, producing the temporal token sequence $\left[\mathrm{CLS}_{\mathrm{temp}};\mathbf{s}_{t-L+1};\ldots;\mathbf{s}_{t}\right]+\mathbf{P}$. The resulting sequence contains $L+1$ tokens: one temporal classification token and $L$ consecutive satellite-state representations.

A second Transformer encoder applies multi-head self-attention across this sequence. This allows changes in communication conditions, hardware behavior, and orbital context to be interpreted relative to preceding observations rather than from the current observation alone. After temporal self-attention, the output corresponding to $\mathrm{CLS}_{\mathrm{temp}}$ is retained as $\mathbf{z}_t$, which summarizes the complete temporal window. Finally, $\mathbf{z}_t$ is passed to a linear classification head that produces the logit vector:

\vspace{-6mm}
\begin{equation}
\hat{\mathbf{y}}_t =
\mathbf{W}_{c}\mathbf{z}_t+\mathbf{b}_{c}.
\label{eq:transformer_classifier}
\end{equation}
\vspace{-3mm}

\noindent The resulting classification corresponds to the final observation $t$ in the temporal window. Overall, Stage 1 models relationships among satellite subsystems within each observation, while Stage 2 models how the resulting fused satellite state evolves across consecutive observations. 

\subsection{Training and Inference}
\label{sec:training_inference}

Both architectures are trained end-to-end for supervised classification for benign and malicious cyberattack detection using the class label associated with the final observation of each temporal window. Given a temporal window $\mathcal{W}_t$ ending at observation $t$, the model produces a logit vector $\hat{\mathbf{y}}_t \in \mathbb{R}^{C}$, where $C$ denotes the number of target classes. The logits are converted into class probabilities using the softmax function:

\begin{equation}
p_{t,c}
=
\frac{\exp(\hat{y}_{t,c})}
{\sum_{j=1}^{C}\exp(\hat{y}_{t,j})},
\label{eq:softmax}
\end{equation}

\noindent where $p_{t,c}$ denotes the predicted probability that the final
observation belongs to class $c$, and $j$ indexes over all $C$ target classes. During training, the model parameters are optimized using weighted multiclass cross-entropy to account for class imbalance:

\vspace{-3mm}
\begin{equation}
\mathcal{L}_{\mathrm{WCE}}
=
-w_{y_t}\log p_{t,y_t},
\label{eq:cross_entropy}
\end{equation}

\noindent where $y_t$ denotes the ground-truth class associated with the final observation of the temporal window and $w_{y_t}$ denotes the corresponding class weight. For each class $c$, the weight is computed as
$w_c=N_{\mathrm{train}}/(C n_c)$, where $N_{\mathrm{train}}$ is the total
number of training targets, $C$ is the number of target classes, and $n_c$ is the number of training targets belonging to class $c$. 

During inference, the models receive only the satellite measurements contained within the current input window; labels from earlier observations are never provided as model inputs. The Subsystem-Fusion MLP processes the temporal measurements through its modality-specific encoders before combining the resulting subsystem representations for classification. The hierarchical Transformer instead constructs a fused satellite-state representation for each observation and then models the sequence of resulting states through its temporal attention stage. In both architectures, the prediction corresponds to the final observation $t$ in the input window. The final predicted class is obtained by selecting the class with the highest predicted probability:

\vspace{-3mm}
\begin{equation}
\hat{c}_t
=
\arg\max_{c \in \{1,\ldots,C\}}
p_{t,c}.
\label{eq:prediction}
\end{equation}

\noindent Therefore, each temporal window produces a single classification for its final observation, while the preceding observations provide contextual information that may improve the resulting decision.

\subsection{Row-Level Variations}
\label{sec:row-level}

In addition to the temporal architectures, we construct matched row-level variations that receive only the current observation $t$. The row-level and temporal models are evaluated on the same target observations, where the row-level model uses only the current observation, while the temporal model also uses preceding observations. This provides a controlled comparison between pointwise and temporal cyberattack detection.

For the Subsystem-Fusion MLP, the row-level variation retains the same modality-specific encoders and fusion mechanism, but removes the temporal window from each subsystem input. Specifically, the temporal model encodes the vectorized modality window $\operatorname{vec}(\mathbf{X}^{(m)}_t)$, while the row-level model receives only the current subsystem feature vector $\mathbf{x}^{(m)}_t$:

\begin{equation}
\mathbf{h}^{(m)}_t =
f_m\left(\mathbf{x}^{(m)}_t\right),
\qquad
m \in \{H,O,R\}.
\label{eq:row_mlp}
\end{equation}

\noindent The resulting hardware, orbital, and RF representations are concatenated using the same fusion operation as the temporal model and passed to the classification head. Therefore, MLP variants differ primarily in whether each subsystem encoder receives a sequence of measurements or a single observation.

For the hierarchical Transformer, the subsystem-level attention stage (Stage 1) is retained and unchanged. The hardware, orbital, and RF measurements from observation $t$ are projected into subsystem tokens and fused through self-attention to obtain the satellite-state representation $\mathbf{s}_t$. The row-level variation then passes $\mathbf{s}_t$ directly to the classification head:

\vspace{-3mm}
\begin{equation}
\hat{\mathbf{y}}_t =
\mathbf{W}_{c}\mathbf{s}_t+\mathbf{b}_{c},
\label{eq:row_transformer}
\end{equation}

\noindent rather than constructing the temporal sequence $[\mathbf{s}_{t-L+1},\ldots,\mathbf{s}_t]$. Consequently, the temporal Transformer, temporal classification token, and temporal positional embeddings are omitted (Stage 2). The subsystem-level Transformer is now responsible for modeling relationships among hardware, orbital, and RF measurements, while no attention is performed across preceding observations. Since the row-level and temporal variations are evaluated on identical target observations, differences in performance reflect the contribution of temporal context rather than differences in the evaluated sample set.

\begin{table*}[t!]
\centering
\caption{Attack-classification performance and inference efficiency under matched row-level and temporal evaluation. Performance metrics are reported as mean $\pm$ standard deviation across runs.}
\label{tab:main_results}
\footnotesize
\setlength{\tabcolsep}{3.6pt}
\begin{tabular}{@{}llccccr@{}}
\toprule
\textbf{Input} &
\textbf{Method} &
\textbf{Acc.} &
\textbf{Prec.} &
\textbf{Rec.} &
\textbf{F1} &
\textbf{Infer. (ms/sample)} \\
\midrule

\multirow{5}{*}{Row}
& Random Forest \cite{azar2023deep}
    & $87.34{\scriptstyle\pm0.33}$
    & $78.08{\scriptstyle\pm0.68}$
    & $75.73{\scriptstyle\pm0.22}$
    & $76.08{\scriptstyle\pm0.42}$
    & $0.246$ \\

& XGBoost \cite{hamidou2025enhancing}
    & $76.64{\scriptstyle\pm0.18}$
    & $61.32{\scriptstyle\pm0.31}$
    & $73.58{\scriptstyle\pm0.11}$
    & $62.39{\scriptstyle\pm0.24}$
    & $0.011$ \\

& Monolithic MLP \cite{cherfi2025mlp}
    & $90.18{\scriptstyle\pm0.74}$
    & $83.90{\scriptstyle\pm2.17}$
    & $76.32{\scriptstyle\pm3.11}$
    & $76.51{\scriptstyle\pm3.49}$
    & $\mathbf{0.008}$ \\

\cmidrule(lr){2-7}

& Subsystem-Fusion MLP
    & $\mathbf{91.35{\scriptstyle\pm0.10}}$
    & $\mathbf{94.44{\scriptstyle\pm0.59}}$
    & $\mathbf{78.52{\scriptstyle\pm0.91}}$
    & $\mathbf{84.65{\scriptstyle\pm0.92}}$
    & $0.144$ \\

& Transformer
    & $91.18{\scriptstyle\pm0.08}$
    & $90.67{\scriptstyle\pm2.26}$
    & $78.71{\scriptstyle\pm0.33}$
    & $82.09{\scriptstyle\pm1.02}$
    & $0.030$ \\

\midrule[\heavyrulewidth]

\multirow{5}{*}{Temporal}
& Random Forest-T \cite{azar2023deep}
    & $83.82{\scriptstyle\pm0.15}$
    & $72.66{\scriptstyle\pm0.08}$
    & $71.41{\scriptstyle\pm0.63}$
    & $71.15{\scriptstyle\pm0.40}$
    & $0.419$ \\

& XGBoost-T \cite{hamidou2025enhancing}
    & $75.50{\scriptstyle\pm0.21}$
    & $63.87{\scriptstyle\pm0.10}$
    & $74.46{\scriptstyle\pm0.04}$
    & $65.30{\scriptstyle\pm0.15}$
    & $0.012$ \\

& Monolithic MLP-T \cite{cherfi2025mlp}
    & $89.77{\scriptstyle\pm0.50}$
    & $80.24{\scriptstyle\pm2.64}$
    & $76.38{\scriptstyle\pm2.14}$
    & $74.90{\scriptstyle\pm2.01}$
    & $\mathbf{0.009}$ \\

\cmidrule(lr){2-7}

& Subsystem-Fusion MLP-T
    & $90.13{\scriptstyle\pm0.44}$
    & $83.46{\scriptstyle\pm0.11}$
    & $75.98{\scriptstyle\pm2.25}$
    & $75.66{\scriptstyle\pm2.20}$
    & $0.132$ \\

& Transformer-T
    & $\mathbf{91.66{\scriptstyle\pm0.02}}$
    & $\mathbf{95.50{\scriptstyle\pm0.21}}$
    & $\mathbf{81.21{\scriptstyle\pm0.13}}$
    & $\mathbf{85.63{\scriptstyle\pm0.14}}$
    & $0.041$ \\

\bottomrule
\end{tabular}
\end{table*}

\section{Experimental Results}
\label{sec:results}

\noindent \textbf{Baselines and Metrics.} We compare the proposed structured architectures against representative learning-based approaches commonly used for satellite and cyberattack detection \cite{azar2023deep,hamidou2025enhancing, cherfi2025mlp}. Specifically, we evaluate Random Forest and XGBoost as conventional machine-learning baselines and implement a monolithic MLP as a deep-learning baseline. The monolithic MLP processes all hardware, orbital, and RF measurements jointly \textit{without} preserving the subsystem-specific organization used by the proposed models. To provide matched temporal comparisons, we also construct temporal variants of each baseline. For Random Forest, XGBoost, and the monolithic MLP, the measurements from all $L$ observations in a temporal window are ordered chronologically and flattened into a single feature vector before being supplied to the model. These temporal variants receive the same observation history and predict the same final-observation targets as the proposed temporal architectures, but do not explicitly model subsystem or sequential structure through attention.

All methods are evaluated using the same data partitions and matched prediction targets. We report accuracy, macro precision, macro recall, and macro F1 across five independent runs with different random seeds. Given the imbalanced class distribution, macro F1 is used as the primary evaluation metric since it assigns equal importance to each target class regardless of its frequency in the dataset.

\noindent \textbf{Implementation Details.}
The neural models were implemented in PyTorch and trained using the MPS backend and conducted on an Apple MacBook Pro equipped with an M5 Pro processor. Neural models were trained for a maximum of 10 epochs with a batch size of 64 using AdamW with a learning rate of $1\times10^{-3}$ and weight decay of $1\times10^{-4}$. Weighted cross-entropy loss was used to address class imbalance. The Transformer uses an embedding dimension of 64, four attention heads, and a feed-forward dimension of 128.

\subsection{Main Experimental Results}
Table \ref{tab:main_results} presents the primary attack-classification results under matched row-level and temporal evaluation. This experiment evaluates each model using the same data partitions and prediction targets \textit{without} holding out an entire satellite for testing. In the row-level setting, each model predicts the class of an observation using only the measurements available at that observation. In the temporal setting, the same target observations are evaluated using an eight-observation window with a stride of one, allowing the models to incorporate the previous satellite measurements when predicting the class of the final observation. Furthermore, inference latency is also reported for each predicted sample.

\begin{figure}[t]
    \centering
    \includegraphics[width=0.7\columnwidth]{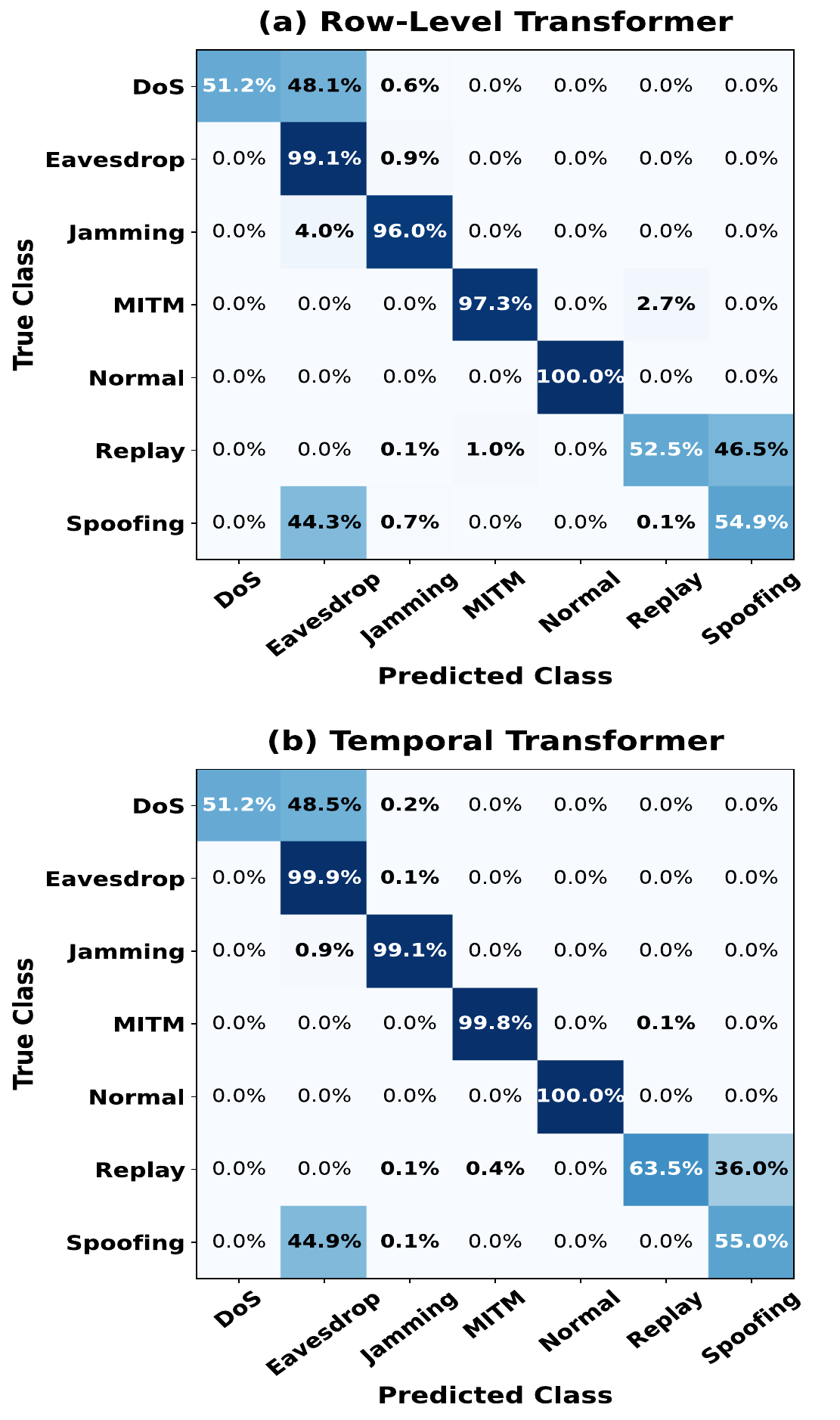}
    \caption{\textbf{Confusion matrices for the (a) row-level and (b) temporal Transformer across five runs. Values represent the percentage of each true class assigned to each predicted class.}}
    \label{fig:confusion}
    \vspace{-5mm}
\end{figure}

Among the row-level methods, the proposed Subsystem-Fusion MLP achieves the highest macro F1 at $84.65\%$, outperforming the row-level Transformer ($82.09\%$) and monolithic MLP ($76.51\%$). This suggests that preserving the subsystem structure of hardware, orbital, and RF can improve the quality of learned representations over a single unstructured input. Under the temporal evaluation experiments, the hierarchical Transformer performs best overall, reaching $91.66\%$ accuracy and $85.63\%$ macro F1. Relative to its row-level counterpart, it improves macro precision by $4.8\%$ and macro recall by $2.5\%$, indicating fewer false-positive predictions relative to true positives and fewer missed class instances.

In contrast, temporal context does not consistently improve the other models. Random Forest-T, monolithic MLP-T, and Subsystem-Fusion MLP-T all achieve lower macro F1 than their row-level counterparts. These models receive preceding observations as a fixed input representation, while the hierarchical Transformer explicitly models relationships across satellite states using temporal self-attention. This suggests that temporal information is most useful when the architecture can explicitly model relationships across observations. Inference latency further highlights the performance-efficiency trade-off. Although the monolithic MLP and XGBoost are fastest, Transformer-T requires only $0.041$ ms per prediction, compared with $0.132$ ms for Subsystem-Fusion MLP-T and $0.419$ ms for Random Forest-T. Therefore, the temporal Transformer achieves the strongest classification performance while maintaining relatively low inference latency.

To further examine class-specific performance, Fig. \ref{fig:confusion} presents the mean normalized confusion matrices for the row-level and temporal Transformer models. Temporal modeling provides the clearest improvements for Replay, Jamming, and MITM attack detection. Replay increases from $52.5\%$ to $63.5\%$, primarily through a reduction in confusion with Spoofing from $46.5\%$ to $36.0\%$. Jamming improves from $96.0\%$ to $99.1\%$, while MITM increases from $97.3\%$ to $99.8\%$. In contrast, DoS remains difficult for both models, with approximately half of its observations classified as Eavesdropping, while Spoofing also remains frequently confused with Eavesdropping. These results indicate that the overall temporal improvement is concentrated in specific attack classes rather than being uniformly distributed across all categories.

\begin{table}[t!]
\centering
\caption{Cross-satellite generalization performance. Source $\rightarrow$ Target denotes the satellite used for training and the unseen satellite used for testing, respectively. Results are reported as macro F1
mean $\pm$ standard deviation across runs.}
\label{tab:cross_satellite}
\footnotesize
\setlength{\tabcolsep}{2.2pt}

\begin{tabular}{@{}llcc@{}}
\toprule
\textbf{Input} &
\textbf{Method} &
\textbf{\shortstack{Sat.~1 $\rightarrow$ Sat.~2}} &
\textbf{\shortstack{Sat.~2 $\rightarrow$ Sat.~1}} \\
\midrule

\multirow{5}{*}{Row}
& Random Forest \cite{udurume2024comparative}
    & $78.78{\scriptstyle\pm1.20}$
    & $\mathbf{88.78{\scriptstyle\pm0.78}}$ \\

& XGBoost \cite{hamidou2025enhancing}
    & $80.08{\scriptstyle\pm0.22}$
    & $79.67{\scriptstyle\pm1.52}$ \\

& Monolithic MLP
    & $78.92{\scriptstyle\pm5.79}$
    & $22.26{\scriptstyle\pm10.67}$ \\
    
\cmidrule(lr){2-4}

& Subsystem-Fusion MLP
    & $\mathbf{80.19{\scriptstyle\pm4.01}}$
    & $87.29{\scriptstyle\pm5.02}$ \\

& Transformer
    & $77.63{\scriptstyle\pm6.68}$
    & $88.33{\scriptstyle\pm3.86}$ \\

\midrule[\heavyrulewidth]

\multirow{5}{*}{Temporal}
& Random Forest-T
    & $70.06{\scriptstyle\pm1.05}$
    & $81.55{\scriptstyle\pm0.64}$ \\

& XGBoost-T
    & $80.95{\scriptstyle\pm0.26}$
    & $84.69{\scriptstyle\pm1.48}$ \\

& Monolithic MLP-T
    & $61.23{\scriptstyle\pm4.74}$
    & $14.74{\scriptstyle\pm4.44}$ \\
    
\cmidrule(lr){2-4}

& Subsystem-Fusion MLP-T
    & $\mathbf{88.47{\scriptstyle\pm0.63}}$
    & $85.86{\scriptstyle\pm3.00}$ \\

& Transformer-T
    & $76.53{\scriptstyle\pm5.38}$
    & $\mathbf{90.06{\scriptstyle\pm1.23}}$ \\

\bottomrule
\end{tabular}
\end{table}

\subsection{Cross-Satellite Generalization}
Table \ref{tab:cross_satellite} reports macro F1 when each model is trained and validated on one satellite and evaluated on the other as an unseen target satellite. This setting provides a more strict generalization test than the primary evaluation, preventing the target satellite from contributing any observations during model development. In addition to measuring transfer between the two satellites represented in UNSW-IoTSAT, the experiment provides an initial indication of how well learned attack representations may transfer to previously unseen satellites. Although this setting does not fully reproduce the variability of an operational LEO constellation, it offers a useful experimental setup for evaluating whether a detector relies primarily on characteristics of the source satellite or captures cyberattack behavior that remains informative across satellite systems.

Cross-satellite performance varies considerably across both model architecture and transfer direction. For Satellite 1 $\rightarrow$ Satellite 2, the temporal Subsystem-Fusion MLP achieves the highest macro F1 of $88.47\%$, substantially improving over its row-level counterpart at $80.19\%$. For Satellite 2 $\rightarrow$ Satellite 1, the temporal Transformer performs best, reaching $90.06\%$ macro F1. In contrast, the monolithic MLP exhibits substantial degradation in both directions, particularly for Satellite 2 $\rightarrow$ Satellite 1, where macro F1 falls to $22.26\%$ for the row-level model and $14.74\%$ for its temporal variant. The strongest cross-satellite results are obtained by the structured temporal models, although transfer performance remains sensitive to the source-target direction.

\subsection{Modality-Ablation}
Table \ref{tab:modality_ablation} reports macro F1 for different combinations of hardware, orbital, and RF inputs under the temporal evaluation setting. This experiment evaluates which satellite information sources contribute most to detection performance for the Subsystem-Fusion MLP-T and Transformer-T models.

Across both architectures, RF measurements provide the strongest individual modality, reaching $75.03\%$ macro F1 for the Subsystem-Fusion MLP-T and $82.38\%$ for Transformer-T, while hardware-only and orbit-only inputs perform substantially worse. For Transformer-T, adding hardware to RF increases macro F1 to $85.47\%$, and the full $H+O+\mathrm{RF}$ configuration reaches $85.63\%$, indicating that most of the improvement beyond RF alone comes from the hardware features. The relatively small gain from adding orbital information suggests that orbital features provide limited additional discriminative value once hardware and RF measurements are available. A similar pattern is observed for the Subsystem-Fusion MLP-T, where adding hardware or orbital features to RF does not consistently improve performance. Overall, RF measurements contain the majority of the discriminative information in UNSW-IoTSAT, while hardware and orbital features can provide complementary information for the temporal models.

\begin{table}[t!]
\centering
\caption{Modality ablation under temporal evaluation. All models use an eight-observation window. Results are reported as macro F1 mean $\pm$ standard deviation across runs.}
\label{tab:modality_ablation}
\footnotesize
\setlength{\tabcolsep}{5pt}
\begin{tabular}{@{}lccc@{}}
\toprule
\textbf{Input Modalities} &
\textbf{\#Feat.} &
\textbf{Fusion MLP-T} &
\textbf{Transformer-T} \\
\midrule

Hardware ($H$)
& 12
& $5.62{\scriptstyle\pm1.04}$ 
& $9.43{\scriptstyle\pm0.10}$ \\

Orbit ($O$)
& 9
& $5.78{\scriptstyle\pm0.71}$
& $6.95{\scriptstyle\pm0.24}$\\


$\mathrm{RF}$
& 10
& $75.03{\scriptstyle\pm1.12}$
& $82.38{\scriptstyle\pm4.63}$ \\

$H+\mathrm{RF}$
& 22
& $73.97{\scriptstyle\pm1.02}$
& $85.47{\scriptstyle\pm0.19}$ \\

$O+\mathrm{RF}$
& 19
& $76.46{\scriptstyle\pm3.43}$
& $80.50{\scriptstyle\pm4.54}$\\

\midrule

$H+O+\mathrm{RF}$
& 31
& $75.66{\scriptstyle\pm2.20}$
& $85.63{\scriptstyle\pm0.14}$ \\

\bottomrule
\end{tabular}
\end{table}

\subsection{Effect of Temporal Window Length}
Table~\ref{tab:window_ablation} examines the sensitivity of the hierarchical Transformer to the amount of temporal context using window lengths $L\in\{2,4,8,16\}$. To ensure a controlled comparison, all window lengths are evaluated on the same prediction targets, defined by the observations with sufficient history to construct a complete 16-observation window. Therefore, each model predicts the same target observations while receiving a different amount of preceding context. This matched-endpoint design isolates the effect of temporal history and explains why the $L=8$ result in this ablation differs slightly from the main experiment, where the 8-observation model was evaluated on the larger set of all valid 8-window endpoints. Among the tested settings, $L=8$ achieves the highest macro F1, reaching $84.78\pm1.26$, and also exhibits the lowest variability across runs. Although the longer $L=16$ window attains competitive performance, it is both slightly lower in mean macro F1 and substantially less stable than $L=8$. Overall, these findings suggest that the hierarchical Transformer benefits from temporal context, but that the benefit saturates beyond approximately eight observations.

\begin{table}[t!]
\centering
\caption{Effect of temporal window length on Transformer-T performance.}
\label{tab:window_ablation}
\footnotesize
\setlength{\tabcolsep}{8pt}

\begin{tabular}{@{}cc@{}}
\toprule
\textbf{Window Length ($L$)} & \textbf{Macro F1} \\
\midrule
2  & $79.35{\scriptstyle\pm6.41}$ \\
4  & $78.22{\scriptstyle\pm5.15}$ \\
8  & $84.78{\scriptstyle\pm1.26}$ \\
16 & $83.81{\scriptstyle\pm4.67}$ \\
\bottomrule
\end{tabular}
\end{table}

\subsection{Effect of Leakage-Resistant Data Partitioning}
Table \ref{tab:leakage_ablation} evaluates the effect of data partitioning on row-level cyberattack classification. We hypothesize that consecutive
observations generated during the same cyberattack occurrence may be highly dependent, such that randomly distributing these observations across training and testing can produce overly optimistic estimates of generalization. To evaluate this effect, we compare the instance-held-out partitioning strategy used throughout the primary experiments with a conventional random-row split. The random-row protocol preserves the same per-class training and test sample counts as the instance-held-out protocol, but does not preserve attack-instance boundaries. All other preprocessing procedures, model configurations, and evaluation settings are held constant.

As predicted, both architectures achieve substantially higher performance under random-row partitioning. The Subsystem-Fusion MLP increases from $84.65\%$ to $92.44\%$ macro F1, while the Transformer increases from $82.09\%$ to $92.26\%$. Accuracy also increases by $4.23\%$ for the Subsystem-Fusion MLP and $4.39\%$ for the Transformer. These results demonstrate that the evaluation partitioning strategy has a substantial effect on detection performance. When observations from the same attack occurrences can appear across training and testing, the resulting performance may overestimate generalization to previously unseen attack instances. This finding supports the use of instance-held-out partitioning for the primary experiments in this study.

\section{Discussion \& Limitations}
The experimental results highlight an important relationship between model architecture and the structure of the available satellite information. At the row level, the Subsystem-Fusion MLP outperforms the Transformer, suggesting that when only a single observation is available, separately encoding subsystem measurements before fusion may be sufficient to capture much of the useful information without requiring attention-based modeling. However, this pattern changes when temporal context is introduced. Providing previous observations does not consistently improve the Random Forest, monolithic MLP, or Subsystem-Fusion MLP, while the hierarchical Transformer improves over its row-level counterpart. These findings suggest that the value of temporal information depends on access to preceding observations and also how the relationships across those observations are modeled. In particular, temporal self-attention in the transformer provides an explicit mechanism for relating representations across the observation window, which may help identify attack behavior that develops over time.

\newcommand{\gain}[1]{\textcolor{blue}{\textbf{+#1}}}

\begin{table}[t]
\centering
\caption{Effect of data partitioning on row-level attack-classification
performance. $\Delta$ denotes the percentage point increase under random-row partitioning.}
\label{tab:leakage_ablation}
\footnotesize
\setlength{\tabcolsep}{3.5pt}

\begin{tabular}{@{}llccc@{}}
\toprule
\textbf{Method} &
\textbf{Metric} &
\textbf{\shortstack{Instance-\\Held-Out}} &
\textbf{Random-Row} &
\textbf{$\Delta$} \\
\midrule

\multirow{2}{*}{Subsystem-Fusion MLP}
& Acc.
& $91.35{\scriptstyle\pm0.10}$
& $95.58{\scriptstyle\pm0.13}$
& \gain{4.23\%} \\

& F1
& $84.65{\scriptstyle\pm0.92}$
& $92.44{\scriptstyle\pm0.47}$
& \gain{7.79\%} \\

\midrule

\multirow{2}{*}{Transformer}
& Acc.
& $91.18{\scriptstyle\pm0.08}$
& $95.57{\scriptstyle\pm0.05}$
& \gain{4.39\%} \\

& F1
& $82.09{\scriptstyle\pm1.02}$
& $92.26{\scriptstyle\pm0.46}$
& \gain{10.17\%} \\

\bottomrule
\end{tabular}
\end{table}

The modality-ablation results further show that detection performance is strongly influenced by the availability of RF measurements. RF features provide the majority of the discriminative information in UNSW-IoTSAT, while hardware and orbital measurements provide additional value for the temporal models. This dependence on RF information should be considered when translating these results to operational satellite systems. Encryption of communication payloads would not necessarily prevent the collection of physical- and link-layer measurements such as received signal strength, SNR, bit-error rate, frequency offset, Doppler shift, or synchronization behavior, since these measurements do not require access to the decoded payload. However, future evaluations should consider partially unavailable RF feature sets to determine how detection performance changes when the complete set of communication measurements cannot be observed.

The proposed architectures present different computational and adaptation trade-offs. The Subsystem-Fusion MLP provides a simpler architecture and achieves the strongest row-level performance, while the hierarchical Transformer achieves the strongest overall performance and maintains relatively low inference latency. Although the MLP may be less computationally expensive to train, the Transformer provides a flexible foundation for future adaptation after deployment. Pretrained transformer-based models can support parameter-efficient adaptation methods through the use of low-rank adaptation \cite{hu2022lora} in which only a small portion of the model is updated, creating opportunities for future continual- and few-shot-learning approaches that incorporate newly observed attack behavior \textit{without} the need to retrain the complete network. These capabilities were not evaluated in the present study and remain an important direction for future work, particularly in satellite environments where new attacks and changing operating conditions may be encountered over time.

Finally, the results should be interpreted within the scope of the UNSW-IoTSAT dataset. Although the cross-satellite experiment provides a stronger generalization test by withholding an entire satellite during model development, the dataset contains only two satellite nodes while real-world operational environments found across LEO constellations may contain thousands of satellite nodes. The observed transfer performance should be viewed as evidence of cross-satellite generalization within the UNSW-IoTSAT environment rather than proof of generalization to arbitrary satellite platforms. Evaluation across additional spacecraft, communication systems, and more expansive datasets as they become publicly available will be necessary to determine how well these learned attack representations transfer beyond the environments considered in this study.

\section{Conclusion}
\label{sec:conclusion}

This paper investigated machine- and deep-learning approaches for classifying benign and malicious behavior in LEO satellites using the UNSW-IoTSAT dataset. Through row-level, temporal, leakage-resistant, and cross-satellite evaluation, we examined how model architecture, temporal context, and data partitioning influence cyberattack detection. We introduced structured Subsystem-Fusion MLP and hierarchical multimodal Transformer architectures that preserve hardware, orbital, and RF information, with the temporal Transformer achieving the strongest overall performance. The results further show that random-row partitioning can inflate performance and that strong within-satellite results do not always translate consistently to unseen satellites. Future work will extend this analysis to additional satellite platforms and operational datasets while exploring adaptive learning methods for emerging attack behavior.

\ifCLASSOPTIONcaptionsoff
  \newpage
\fi

\bibliographystyle{IEEEtran}
\bibliography{ref}

\end{document}